\documentclass[prb,twocolumn,superscriptaddress]{revtex4}

\usepackage{amsmath,graphicx,latexsym,bm,subfigure,mathrsfs}
\graphicspath{{./Figures/}}

\newcommand{\K}{\ensuremath{\,\mbox{K}}}
\newcommand{\celsius}{\ensuremath{\,{}^\circ}\!C}
\newcommand{\cm}{\ensuremath{\,\mbox{cm}^{-1}}}

\begin{document}

\title{First-principles design and subsequent synthesis of a material to
search for the permanent electric dipole moment of the electron}

\author{K.~Z.~Rushchanskii}
\affiliation{Institut f\"{u}r Festk\"{o}rperforschung, Forschungszentrum J\"{u}lich GmbH 52425 J\"{u}lich and JARA-FIT, Germany}

\author{ S.~Kamba}
\affiliation{Institute of Physics ASCR, Na Slovance~2, 182 21 Prague~8, Czech Republic}

\author{ V.~Goian}
\affiliation{Institute of Physics ASCR, Na Slovance~2, 182 21 Prague~8, Czech Republic}

\author{ P.~Van\v{e}k}
\affiliation{Institute of Physics ASCR, Na Slovance~2, 182 21 Prague~8, Czech Republic}

\author{ M.~Savinov}
\affiliation{Institute of Physics ASCR, Na Slovance~2, 182 21 Prague~8, Czech Republic}

\author{ J.~Prokle\v{s}ka}
\affiliation{Charles University, Faculty of Mathematics and Physics, Department of Condensed Matter Physics, Ke Karlovu 5, 121 16 Prague 2, Czech Republic}

\author{ D.~Nuzhnyy}
\affiliation{Institute of Physics ASCR, Na Slovance~2, 182 21 Prague~8, Czech Republic}

\author{ K.~Kn\'{i}\v{z}ek}
\affiliation{Institute of Physics ASCR, Na Slovance~2, 182 21 Prague~8, Czech Republic}

\author{ F.~Laufek}
\affiliation{Czech Geological Survey, Geologick\'{a} 6, 152 00 Prague~5, Czech Republic}

\author{S.~Eckel}
\affiliation{Yale University, Department of Physics, P.O. Box 208120, New Haven, CT 06520-8120}

\author{S.~K.~Lamoreaux}
\affiliation{Yale University, Department of Physics, P.O. Box 208120, New Haven, CT 06520-8120}

\author{A.~O.~Sushkov}
\affiliation{Yale University, Department of Physics, P.O. Box 208120, New Haven, CT 06520-8120}

\author{M.~Le\v{z}ai\'{c}}
\affiliation{Institut f\"{u}r Festk\"{o}rperforschung, Forschungszentrum J\"{u}lich GmbH 52425 J\"{u}lich and JARA-FIT, Germany}

\author{N.~A.~Spaldin}
\affiliation{Materials Department, University of California, Santa Barbara,
             CA 93106-5050, USA}

\date{\today}

\begin{abstract}
We describe the first-principles design and subsequent synthesis of a new
material with the specific functionalities required for a solid-state-based
search for the permanent electric dipole
moment of the electron. We show computationally that perovskite-structure
europium barium titanate should exhibit the required large and pressure-dependent
ferroelectric polarization, local magnetic moments, and absence of magnetic
ordering at liquid helium temperature. Subsequent synthesis and characterization
of Eu$_{0.5}$Ba$_{0.5}$TiO$_3$ ceramics confirm the predicted desirable properties.
\end{abstract}

\maketitle


\begin{figure}
\begin{center}
\resizebox{0.7\columnwidth}{!}{\includegraphics{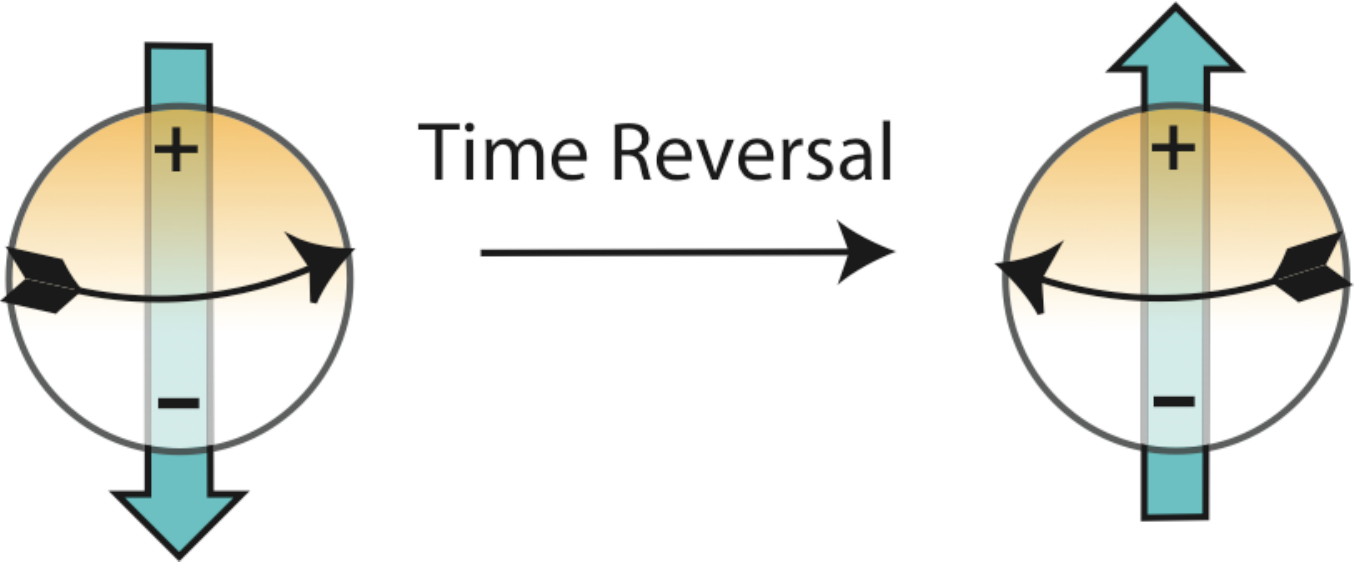}} \\
\end{center}
\caption{Illustration that an electron with an electric dipole moment violates
time-reversal symmetry. Both the electric dipole moment (+ and - symbols; orange shading)
and magnetic moment (blue arrow) of the electron lie along the same axis as 
the electron spin (black arrow). The operation of time reversal reverses the 
magnetic moment but does not affect the electric dipole moment; therefore an 
electron with a non-zero electric dipole moment violates time-reversal symmetry. }
\label{PT_cartoon}
\end{figure}

The Standard Model of particle physics incorporates the breaking of the
discrete symmetries of parity ($P$) and the combined charge conjugation and
parity ($CP$). It is thought however, that the $CP$-violation within the
framework of the Standard Model is insufficient to explain the observed
matter-antimatter asymmetry of the Universe~\cite{Trodden1999}, therefore a
so far unknown source of $CP$-violation likely exists in nature. The
existence of a non-zero permanent electric dipole moment (EDM) of a
particle, such as an electron, neutron, or atom, would violate time reversal
($T$) symmetry (Fig.~\ref{PT_cartoon}) and therefore imply $CP$-violation 
through the $CPT$
theorem~\cite{Khriplovich1997}. In the Standard Model these EDMs are
strongly suppressed, the theoretical predictions lying many orders of
magnitude below the current experimental limits.
However, many theories beyond the Standard Model, such as supersymmetry,
contain a number of $CP$-violating phases that lead to EDM predictions
within experimental reach~\cite{Bernreuther1991}. Searching for EDMs
therefore constitutes a background-free method of probing the $CP$-violating
physics beyond the Standard Model.

\begin{figure}
\begin{center}
\resizebox{\columnwidth}{!}{\includegraphics{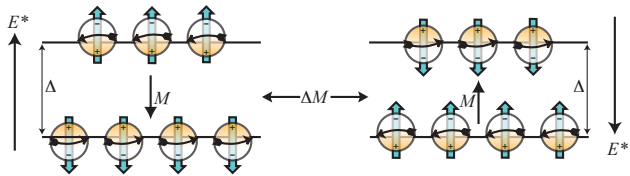}} \\
\end{center}
\caption{Schematic of the physics underlying the experiment to search for the electron
EDM. The energy of electrons with EDMs parallel to the effective electric field $E^*$ 
is lower than that for electrons with anti-parallel EDMs by an 
amount $\Delta = E^* \cdot d_e$. As a 
result there is a population imbalance (exaggerated for clarity in the figure), and, 
since the magnetic moments are 
aligned along the EDM directions, a corresponding net magnetization, $M$. When
the electric field is reversed there is a magnetization reversal, $\Delta M$,
which can be detected using a sensitive magnetometer.}
\label{Zeeman}
\end{figure}

A number of experimental EDM searches are currently under way or are being
developed -- systems studied in these experiments include
diatomic molecules~\cite{Hudson2002,Kawall2004}, diamagnetic
atoms~\cite{Griffith2009,Guest2007,Tardiff2007}, molecular
ions~\cite{Stutz2004}, cold atoms~\cite{Weiss2003},
neutrons~\cite{Baker2006}, liquids~\cite{Ledbetter2005}, and
solids~\cite{Heidenreich2005,Bouchard2008} -- 
with one of the most promising novel techniques being electric-field-correlated
magnetization measurements in solids~\cite{Shapiro1968,Lamoreaux2002,Budker2006}. 
This technique rests on the fact
that, since spin is the only intrinsic vector associated with the electron, a non-vanishing electron EDM
is either parallel or antiparallel to its spin and hence 
its magnetic moment. As a result, when an electric field, which lifts the degeneracy between
electrons with EDMs parallel and antiparallel to it, is applied to a sample, the associated 
imbalance of electron populations generates a 
magnetization (Fig.~\ref{Zeeman}). The orientation of the magnetization is reversed 
when the electric field direction is switched; in our proposed experiment we will 
monitor this change in sample magnetization using
a SQUID magnetometer~\cite{Sushkov2009,Sushkov2010}.
Such {\it magnetoelectric responses} in materials with permanent {\it macroscopic}
magnetizations and polarizations are of great current interest in the 
materials science community because of their potential for enabling novel 
devices that tune and control magnetism using electric fields\cite{Spaldin/Ramesh:2008}. 

Since the experiment aims to detect the intrinsic magnetoelectric response associated with the
tiny electric dipole moment of the electron, the design constraints on the material are
stringent. First, the solid must contain magnetic ions with unpaired spins, since 
the equal and opposite spins of paired electrons have corresponding equal and opposite
EDMs and contribute no effect. Second, it must be engineered such that the {\it conventional} 
linear magnetoelectric tensor is zero; our approach to achieving this is to use a paramagnet
in which the conventional effect is forbidden by time-reversal symmetry\cite{Fiebig:2005}.
To reach the required sensitivity, a high atomic density of magnetic
ions ($n\approx 10^{22}$~cm$^{-3}$) is needed, and these magnetic ions must reside
at sites with broken inversion symmetry. The energy splitting $\Delta$ shown
in Fig.~\ref{Zeeman} is proportional to the product of the effective electric field
experienced by the electron, $E^*$, and its electric dipole moment, $d_e$. 
The effective electric field, which is equal to the electric field one would have
to apply to a free electron to obtain the same energy splitting, is in turn
determined by the displacement
of the magnetic ion from the center of its coordination polyhedron;
for a detailed derivation see Ref. \cite{Mukhamedjanov2003}.
For example,
in Eu$_{0.5}$Ba$_{0.5}$TiO$_3$ ceramics (see below) with $\sim$1 $\mu$C/cm$^2$ remanent
polarization, the mean displacement of the Eu$^{2+}$ ion with respect to its oxygen cage
is 0.01 \AA\ and this results in an effective electric field of $\sim$10 MV/cm, even
when no external electric field is applied.
We choose a ferroelectric so that it is possible to reverse the direction of the
ionic displacements, and hence of the effective electric field, with a moderate applied
electric field. Finally, the experiment will be performed inside liquid helium,
so the material properties described above must persist at low temperature.
A detailed derivation of the
dependence of the sensitivity on the material parameters is given in 
Ref.~\cite{Sushkov2010}. Note that conventional impurities such as defects or
domain walls are not detrimental to the experiment since they do not violate
time-reversal symmetry. 
In summary, the following material specifications will allow a sensitive EDM
search to be mounted:
(i) The material should be ferroelectric, with a large electric polarization, 
and switchable at liquid He temperature. (ii) There should be a high concentration 
of ions with local magnetic moments that remain paramagnetic at liquid He temperature;
both long-range order and freezing into a glassy state must be avoided.
(iii) The local environment at each magnetic ion should be strongly modified by
the ferroelectric switching, and (iv) the sample should be macroscopic.
With these material properties, and optimal SQUID noise levels, 
the projected experimental sensitivity is 10$^{-28}$ e.cm after ten days 
of averaging\cite{Sushkov2010}. 

No known materials meet all the requirements. Indeed the contra-indication between
ferroelectricity and magnetism has been studied extensively over the last decade
in the context of multiferroics \cite{Hill:2000}, where the goal has been to
achieve simultaneous ferroelectric and ferromagnetic ordering at high temperature.
In spite of extensive efforts,
a room temperature multiferroic with large and robust ferroelectricity
and magnetization at room temperature remains elusive. While the low temperature
constraints imposed here seem at first sight more straightforward, avoiding any magnetic
ordering at low temperature, while retaining a high concentration of magnetic
ions poses a similarly demanding challenge. In addition the problem of ferroelectric
switchability at low temperature is challenging, since coercivities tend to
increase as temperature is lowered \cite{Merz:1951}.

We proceed by proposing a trial compound and calculating its properties using
density functional theory to determine whether an experimental synthesis should 
be motivated. 
We choose an alloy of europium titanate, EuTiO$_3$ and barium titanate, 
BaTiO$_3$, with motivation as follows:
To incorporate
magnetism we require unfilled orbital manifolds of localized electrons; to avoid magnetic
ordering the exchange interactions should be small. Therefore the tightly bound $4f$ electrons
are likely to be the best choice. For conventional ferroelectricity
we require
transition metal ions with empty $d$ orbitals to allow for good hybridization with
coordinating anions on off-centering \cite{Rondinelli/Eidelson/Spaldin:2009}.
(Note that while here we use a conventional ferroelectric mechanism, many 
alternative routes to ferroelectricity that are compatible with magnetism 
-- and which could form a basis for future explorations -- have been recently 
identified; for a review see Ref.~\onlinecite{Ramesh/Spaldin:2007}).
Both EuTiO$_3$ and BaTiO$_3$ form in the ABO$_3$ perovskite structure, with divalent Eu$^{2+}$
or Ba$^{2+}$ on the A site, and formally $d^0$ Ti$^{4+}$ on the B site. BaTiO$_3$ is a
prototypical ferroelectric with a large room temperature polarization of 25
$\mu$C/cm$^2$.\cite{Wemple:1968} In the cubic paraelectric phase its
lattice constant is 3.996 \AA\ \cite{Miyake/Ueda:1947}. 
The Ba$^{2+}$ ion has an inert gas electron configuration and hence zero
magnetic moment.

The lattice parameter of EuTiO$_3$ is 
3.905 \AA\ \cite{Katsufuji/Takagi:2001}, notably smaller than that of BaTiO$_3$.
It is not ferroelectric, but has a large dielectric constant ($\epsilon \approx 400$)
at low temperature, indicative of proximity to a ferroelectric phase transition; indeed
it has recently been reported to be a quantum paraelectric\cite{Katsufuji/Takagi:2001,kamba:2007}.
First-principles electronic structure calculations have shown that ferroelectricity
should be induced along the elongation direction by either compressive or tensile strain
\cite{Fennie/Rabe:2006}.
The Eu$^{2+}$ ion has seven unpaired localized $4f$
electrons resulting in a large spin magnetization of 7 $\mu_B$, and EuTiO$_3$ is an
antiferromagnet with $G$-type ordering at a low N\'{e}el temperature of $\sim$5.3\,K
\cite{McGuire_et_al:1966,Chien/DeBenedetti/Barros:1974}.
(Independently of the study presented here, EuTiO$_3$ is of considerable current 
interest because its dielectric response is strongly affected by the magnetic ordering
\cite{Katsufuji/Takagi:2001,kamba:2007} and because of its unusual third order magnetoelectric
response \cite{Shvartsman_et_al:2010}. These behaviors indicate
coupling between the magnetic and dielectric orders caused by sensitivity of the
polar soft mode to the magnetic ordering \cite{Fennie/Rabe:2006,Goian:2009}.)

Our hypothesis is that by alloying Ba on the A-site of
EuTiO$_3$, the magnetic ordering temperature will be suppressed through dilution, and the
tendency to ferroelectricity will be increased through the expansion of the lattice
constant. Our hope is to identify an alloying range in which the magnetic ordering 
temperature is sufficiently low while the ferroelectric polarization and the concentration 
of magnetic ions remain sufficiently large. In addition, we expect that the polarization
will be sensitive to the lattice constant, allowing its magnitude and consequently
the coercivity, to be reduced with pressure.

\section{First-Principles Calculations}

\begin{figure}
\begin{center}
\resizebox{0.9\columnwidth}{!}{\includegraphics{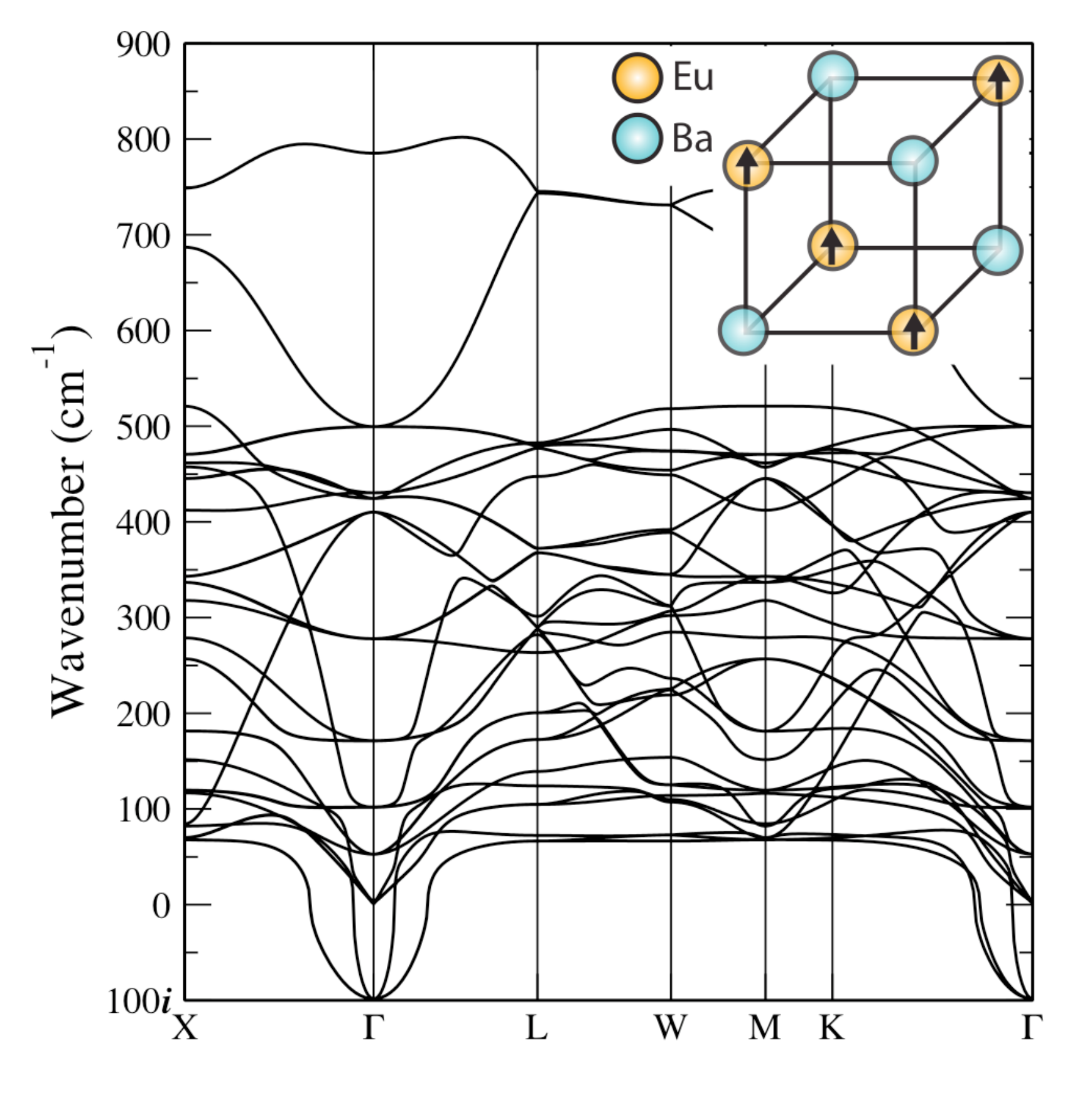} }
\end{center}
\caption{
Calculated phonon dispersion of ferromagnetic Eu$_{0.5}$Ba$_{0.5}$TiO$_3$
in its high symmetry reference structure with pseudo-cubic lattice constant $a_0=3.95$ \AA. 
The imaginary frequency polar phonon at $\Gamma$ indicates a structural instability 
to a ferroelectric phase.
The inset shows the supercell of the ferromagnetic Eu$_{0.5}$Ba$_{0.5}$TiO$_3$ ordered alloy used 
in our calculations. The Ti and O ions are omitted for clarity; arrows represent the Eu magnetic
moments. 
\label{th_phonons}}
\end{figure}

Taking the 50/50 (Eu,Ba)TiO$_3$ ordered alloy as our starting point
(Fig.~\ref{th_phonons} inset), 
we next calculate its
properties using first-principles. For details of the computations
see the Methods section. 

We began by calculating the phonon dispersion for the high symmetry, cubic perovskite
reference structure at a lattice constant of 3.95 \AA\ (chosen, somewhat arbitrarily,
for this first step because it is the average 
of the experimental BaTiO$_3$ and EuTiO$_3$ lattice constants), with the magnetic spins
aligned ferromagnetically; our results are shown in Fig.~\ref{th_phonons}, plotted
along the high symmetry lines of the Brillouin zone. 
Importantly we find a polar $\Gamma$-point instability with an imaginary frequency of 
103$i$~cm$^{-1}$
which is dominated by relative oxygen -- Ti/Eu displacements (the eigenmode displacements for
Eu, Ba, Ti, O$_{\parallel}$ and O$_{\perp}$ are 0.234, -0.059, 0.394, -0.360 and -0.303
respectively); such polar instabilities are indicative
of a tendency to ferroelectricity. The zone boundary rotational instabilities that often occur in
perovskite oxides and lead to non-polar, antiferrodistortive ground states are notably absent
(in fact the flat bands at $\sim$60 cm$^{-1}$ are stable rotational vibrations).
Interestingly we find
that the Eu ions have a significant amplitude in the soft-mode eigenvector, in contrast
to the Ba ions both here and in the parent BaTiO$_3$.

Next we performed a structural optimization of both the unit cell shape and the
ionic positions of our Eu$_{0.5}$Ba$_{0.5}$TiO$_3$ alloy with the total volume constrained
to that of the ideal cubic structure studied above (3.95$^3$ \AA\,$^3$ per formula unit).
Our main finding is that the
Eu$_{0.5}$Ba$_{0.5}$TiO$_3$ alloy is polar with large relative displacements of oxygen
and both Ti and Eu relative to the high symmetry reference structure. Using the Berry
phase method we obtain a ferroelectric polarization value of $P = 23$ $\mu$C/cm$^2$. 
Our
calculated ground state is orthorhombic with the polarization oriented along a [011]
direction and lattice parameters $a=3.94$ \AA, $b=5.60$ \AA\ and $c=5.59$ \AA. As
expected from our analysis of the soft mode, the calculated ground state is characterized
by large oxygen -- Ti/Eu displacements, and the absence of rotations or tilts of the
oxygen octahedra. Importantly, the large Eu amplitude in the soft mode manifests as a
large off-centering of the Eu from the center of its oxygen coordination polyhedron in
the ground state structure. The origin of the large Eu displacement lies
in its small ionic radius compared with that of divalent Ba$^{2+}$:
The large coordination cage around the Eu ion which is imposed by the large lattice
constant of the alloy results in under-bonding of the Eu that can be relieved by
off-centering. Indeed, we find that in calculations for fully relaxed single phase
EuTiO$_3$, the oxygen octahedra tilt to reduce the volume of the A site in a similar 
manner to those known to occur in SrTiO$_3$, in which the A cation size is almost 
identical. This Eu off-centering is
desirable for the EDM experiment because the change in local environment at the magnetic
ions on ferroelectric switching determines the sensitivity of the EDM measurement.

\begin{table}[htbp]
\centering
\begin{tabular}{l l c}\hline  
\multicolumn{2}{c}{volume (\AA\,$^3$)}  &  P ($\mu$C/cm$^2$)\\ \hline
61.63          & (constrained)  &  23               \\
62.30          & (experimental) &  28               \\
64.63          & (relaxed)      &  44               \\
\hline\end{tabular}
\caption{Calculated ferroelectric polarizations, P,  of Eu$_{0.5}$Ba$_{0.5}$TiO$_3$ at
three different volumes.}
\label{PversusV}
\end{table}

We note that the magnitude of the polarization is strongly dependent on the volume
used in the calculation (Table~\ref{PversusV}). 
At the experimental volume (reported in the next section),
which is only slightly larger than our constrained volume of $3.95^3$ \AA\,$^3$,
we obtain a polarization of 28 $\mu$C/cm$^2$. 
At full relaxation, where we find a larger volume close to that
of BaTiO$_3$, we obtain a polarization of 44 $\mu$C/cm$^2$, almost certainly a 
substantial over-estimate. 
This volume dependence suggests that the use
of pressure to reduce the lattice parameters and suppress the ferroelectric
polarization could be a viable tool for reducing the coercivity at low temperatures. Indeed
our computations show that, at a pressure corresponding to 2.8 GPa applied to the experimental volume 
the theoretical structure is cubic, with both the polarization and coercive field reduced 
to zero. 

Finally, to investigate the likelihood of magnetic ordering,
we calculated the relative energies of the ferromagnetic state discussed above
and of two antiferromagnetic arrangements: planes
of ferromagnetically ordered spins coupled antiferromagnetically along either the
pseudo-cubic $z$ axis or the $x$ or $y$ axes. 
(Note that these are degenerate in the high-symmetry cubic structure).
For each magnetic arrangement we re-relaxed the lattice parameters and atomic
positions.
As expected for the highly localized Eu $4f$ electrons on their diluted sublattice,
the energy differences between the different configurations are small
-- around 1 meV per 40 atom supercell -- suggesting an absence of magnetic ordering
down to low temperatures. While our calculations find the ferromagnetic state to
be the lowest energy, this is likely a consequence of our A-site ordering
and should not lead us to anticipate ferromagnetism at low temperature
(Note that, after completing our study, we found a report of an early 
effort to synthesize (Eu,Ba)TiO$_3$\cite{Janes/Bodnar/Taylor:1978} in which a
large magnetization, attributed to A-site ordering and ferromagnetism, was
reported. A-site ordering is now known to be difficult to achieve in perovskite-structure
oxides, however, and we find no evidence of it in our samples. Moreover the earlier work
determined a tetragonal crystal structure in contrast to our refined orthorhombic structure.) 

In summary, our predicted properties of the (Eu,Ba)TiO$_3$ alloy -- large ferroelectric
polarization, reducible with pressure, with large Eu displacements, and strongly
suppressed magnetic ordering -- meet the criteria for the electron electric dipole 
moment search and motivate the synthesis and characterization of the compound, 
described next.

\section{Synthesis}
Eu$_{0.5}$Ba$_{0.5}$TiO$_3$ was synthesized by solid-state reaction using mechanochemical
activation before calcination. For details see the Methods section. 
The density of the sintered pellets was 86-88\% of the
theoretical density. X-ray diffraction at room temperature revealed the cubic perovskite
$Pm\bar{3}m$ structure with a=3.9642(1)\,\AA. At 100\,K we obtain an orthorhombic ground
state with space group $Amm2$, in agreement with the GGA$+U$ prediction,  and lattice 
parameters 3.9563(1), 5.6069(2) and 5.5998(2) \AA.

\section{Characterization}
\begin{figure}
\begin{center}
\resizebox{0.9\columnwidth}{!}{\includegraphics{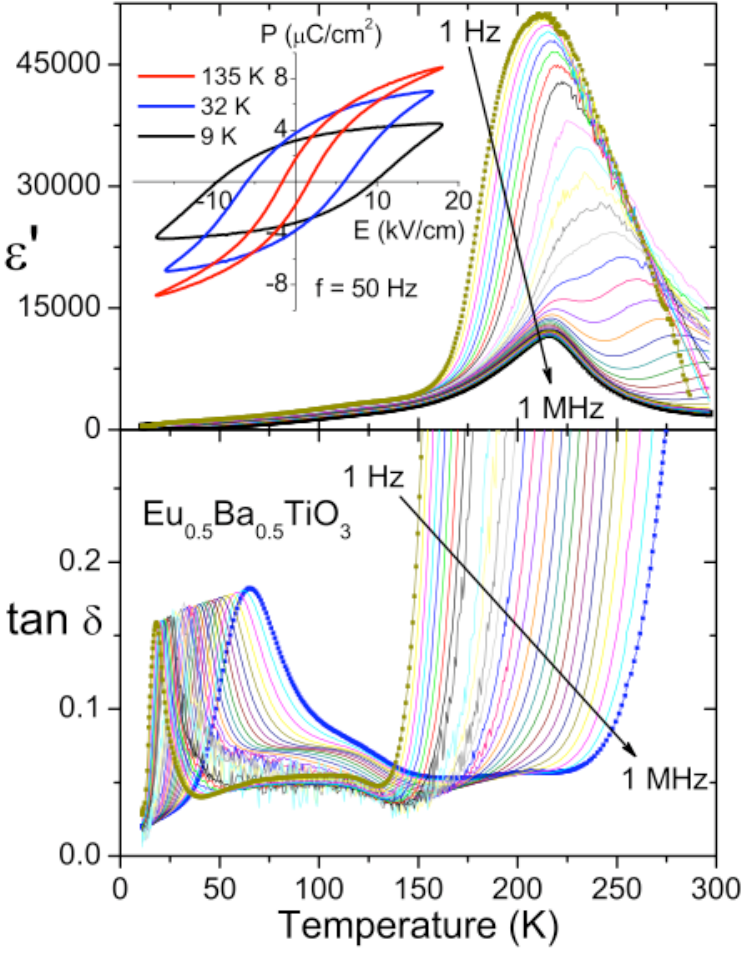} }
\end{center}
\caption{Temperature dependence of permittivity and dielectric loss in
Eu$_{0.5}$Ba$_{0.5}$TiO$_3$ ceramics. The arrows indicate the direction
of increasing frequency and the colors are for clarity to assist the eye
in distinguishing the
lines. The inset shows ferroelectric hysteresis loops measured
at three temperatures and 50\,Hz.} \label{Fig3}
\end{figure}

The final step in our study is the characterization of the samples, to
check that the measured properties are indeed the same as those that we predicted and 
desired.
Figure~\ref{Fig3} shows the temperature dependence of the complex permittivity between
1\,Hz and 1\,MHz, measured using an impedance analyzer ALPHA-AN (Novocontrol). The
low-frequency data below 100\,kHz are affected above 150\K\, by a small defect-induced
conductivity and related Maxwell-Wagner polarization; the high-frequency data clearly
show a maximum in the permittivity near $T_c$=213\K\ indicating the ferroelectric phase transition.
Two regions of dielectric dispersion -- near 100\K\ and below 75\K\ -- are seen in 
tan$\delta(T)$; these could originate from oxygen defects or from ferroelectric domain 
wall motion.

Measurement of the polarization was adversely affected by the sample conductivity above 
150\K, but at lower temperatures good quality ferroelectric hysteresis loops were obtained 
(Fig.~\ref{Fig3}, inset). At 135\K\, we obtain a saturation polarization of $\sim$8 $\mu$C/cm$^2$.
The deviation from the predicted value could be the result of incomplete saturation as
well as the strong volume dependence of the polarization combined with the well-known 
inaccuracies in GGA$+U$ volumes.
As expected, at lower temperatures the coercive field strongly increases, and only
partial polarization switching was possible even with an applied electric field of
18\,kV/cm (at higher electric field dielectric breakdown was imminent). The partial 
switching is responsible for the apparent decrease in saturation polarization below 40\,K.

\begin{figure}[h]
\begin{center}
\resizebox{0.9\columnwidth}{!}{\includegraphics{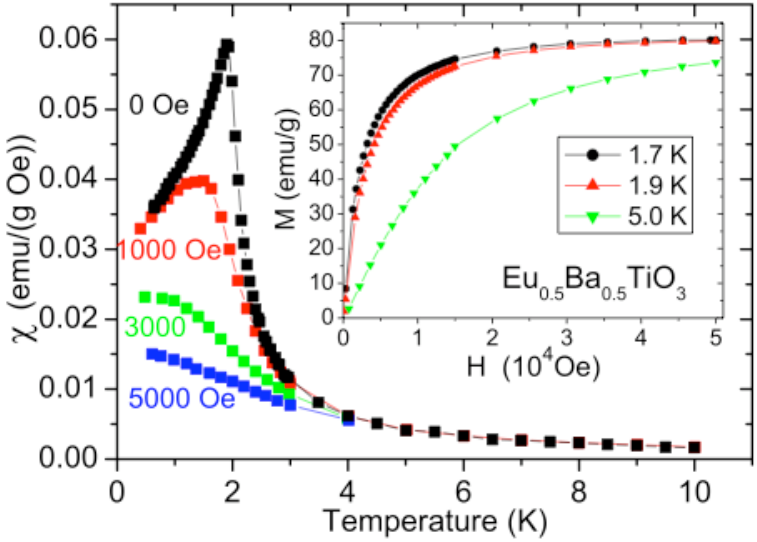} }
  \end{center}
\caption{Temperature dependence of ac magnetic susceptibility, $\chi$, at various static magnetic
fields and frequency of 214\,Hz. Inset shows magnetization curves at various
temperatures. We note that no hysteresis in magnetization was observed.} \label{Fig4}
\end{figure}

Time-domain THz transmission and infrared reflectivity spectra (not shown here) reveal
a softening of the polar phonon from $\sim$40\cm\, at 300\K\ to $\sim$15\cm\, at $T_c$, 
and then its splitting into two components in the ferroelectric phase. Both components
harden on cooling below $T_c$, with the lower frequency component remaining below 20\cm\, 
down to 10\K, and the higher-frequency branch saturating near 70\cm\, at 10\K. This
behavior is reminiscent of the soft-mode behavior in BaTiO$_{3}$\cite{Hlinka:2008}.
However, when we extract the contribution to the static permittivity that comes from 
the polar phonon, we find that it is
considerably smaller than our measured value (Fig.~\ref{Fig3}) indicating an additional
contribution to the dielectric relaxation. Our observations suggest that the phase transition 
is primarily soft-mode driven, but also exhibits some order-disorder character. 

Finally, we measured the magnetic susceptibility $\chi$ at various static magnetic 
fields as a function of temperature down to 0.4\,K. (For details see the Methods section.)
Our results are shown in Fig.~\ref{Fig4}. $\chi(T)$ peaks at $T\sim$1.9\,K indicating an
absence of magnetic ordering above this temperature. 
The $\chi(T)$ data up to 300\,K show Curie-Weiss behavior 
$\chi(T)=\frac{C}{T+\theta}$ with 
$\theta$=-1.63\,K and $C = 0.017$ emuK/(gOe). 
The peak in susceptibility at 1.9K is frequency independent and not influenced by
zero field heating measurements after field cooling, confirming antiferromagnetic
order below $T_N = 1.9$\,K.
As in pure EuTiO$_3$, the $\chi(T)$ peak is suppressed by a
static external magnetic field, indicating stabilization of the paramagnetic phase
\cite{Katsufuji/Takagi:2001}. 
Magnetization curves (Fig.~\ref{Fig4} inset) show saturation above $2\times10^4$ Oe 
at temperatures below $T_{N}$ and slower saturation at 5\,K. No open magnetic hysteresis 
loops were observed.

In summary, we have 
designed a new material -- Eu$_{0.5}$Ba$_{0.5}$TiO$_3$ -- with the properties required
to enable a measurement of the EDM to a higher accuracy than can currently be
realized. Subsequent synthesis of Eu$_{0.5}$Ba$_{0.5}$TiO$_3$ ceramics confirmed
their desirable ferroelectric polarization and absence of magnetic ordering above
1.9\,K. 
The search for the permanent dipole moment of the electron using Eu$_{0.5}$Ba$_{0.5}$TiO$_3$
is now underway. 
Initial measurements have already achieved an EDM upper limit of 5 $\times
10^{-23}$ e.cm, which is within a factor of 10 of the current record with a
solid-state-based EDM search \cite{Heidenreich2005}. 
We are currently studying a number of systematic effects that may mask the
EDM signal. The primary error originates 
from ferroelectric hysteresis-induced heating of the samples during
polarization reversal. 
This heating gives rise to a change in magnetic susceptibility,
which, in a non-zero external magnetic field, leads to an undesirable sample
magnetization response. We are working to control the absolute magnetic
field at the location of the samples to the 0.1 $\mu$G level. 
Our projected sensitivity of 10$^{-28}$ e.cm should then be achievable.

\section{Acknowledgments}
This work was supported by the US National Science Foundation under award number
DMR-0940420 (NAS), by Yale University, by the Czech Science Foundation (Project Nos. 
202/09/0682 and AVOZ10100520) and the
Young Investigators Group Programme of Helmholtz Association, Germany, contract VH-NG-409.
We thank O. Pacherova, R. Krupkova and G. Urbanova for technical assistance and Oleg 
Sushkov for invaluable discussions.

\section{Author contributions }
SKL supervised the EDM measurement effort at Yale. AOS and SE performed the analysis and made
preliminary measurements, showing that these materials could be useful in an EDM experiment.
ML and NAS selected (Eu,Ba)TiO$_3$ as the candidate material according to the experimental
requirements and supervised the ab-initio calculations. KZR performed the ab-initio calculations.
ML, NAS and KZR analysed the ab-initio results and wrote the theoretical component of the paper.
Ceramics were prepared by PV. Crystal structure was determined by KK and FL. Dielectric measurements
were performed by MS. JP investigated magnetic properties of ceramics. VG performed infrared
reflectivity studies. DN investigated THz spectra. SK coordinated all experimental studies and
wrote the synthesis and characterization part of manuscript. NAS coordinated the preparation of
the manuscript.

\section{Methods}

\subsection{Computational details}
We performed first-principles density-functional calculations within the spin-polarized generalized
gradient approximation (GGA) \cite{PBE:1996}. The strong on-site correlations of the Eu $4f$
electrons were treated using the GGA+$U$ method \cite{Anisimov/Aryasetiawan/Liechtenstein:1997} with the double
counting treated within the Dudarev approach \cite{Dudarev_et_al:1998} and parameters
$U=5.7$~eV and $J=1.0$~eV. For structural relaxation and lattice
dynamics we used the Vienna \textit{Ab Initio} Simulation Package (VASP)
\cite{VASP_Kresse:1996} with the default
projector augmented-wave (PAW) potentials~\cite{Bloechl:1994} (valence-electron configurations
Eu: $5s^2 5p^6 4f^{7}6s^{2}$, Ba: $5s^{2}5p^{6}6s^{2}$, Ti: $3s^{2}3p^{6}3d^{2}4s^{2}$ and O: $2s^{2}2p^{4}$.)
Spin-orbit interaction was not included.

The 50/50 (Eu,Ba)TiO$_3$ alloy was represented by an ordered A-site structure with the
Eu and Ba ions alternating in a checkerboard pattern (Fig.~\ref{th_phonons}, inset). Structural
relaxations and total
energy calculations were performed for a 40-atom supercell (consisting of two 5-atom perovskite
unit cells in each cartesian direction) using a $4\times4\times4$ $\Gamma$-centered $k$-point
mesh and a plane-wave cutoff of 500 eV.
Ferroelectric polarizations and Born effective charges were calculated using the Berry phase method
\cite{King-Smith:1993}. 
Lattice instabilities were investigated in the frozen-phonon scheme \cite{Kunc:1982, Alfe:2009}
for an 80 atom supercell using a $\Gamma$-centered $2\times2\times2$ $k$-point mesh
and 0.0056~\AA\ atomic displacements to extract the Hellman-Feynman forces.

\subsection{Synthesis}
Eu$_2$O$_3$, TiO$_2$ (anatase) and BaTiO$_3$ powders (all
from Sigma-Aldrich) were mixed in stoichiometric ratio then milled intensively in a 
planetary ball micro mill Fritsch Pulverisette 7 for 120\,min. in a dry environment 
followed by 
20 min. in suspension with n-heptane. ZrO$_2$ grinding bowls (25\,ml) and balls (12\,mm
diameter, acceleration 14\,g) were used. The suspension was dried under an IR lamp and the 
dried
powder was pressed in a uniaxial press (330\,MPa, 3\,min.) into 13\,mm diameter
pellets. The
pellets were calcined in pure H$_2$ atmosphere at 1200\celsius\ for 24\,hr (to reduce
Eu$^{3+}$ to Eu$^{2+}$), then milled and pressed by the same procedure as above and 
sintered at 1300\celsius\ for 24\,hr in Ar\,+\,10\%\,H$_2$ atmosphere. Note that pure
H$_2$ can not be used for sintering without adversely increasing the conductivity of
the sample. 

\subsection{Characterization}
Magnetic susceptibility was measured using a Quantum Design PPMS9 and a 
He$^3$ insert equipped with a home-made induction coil that allows measurement 
of ac magnetic susceptibility, $\chi$ from 0.1 to 214\,Hz.

\end{document}